\documentclass{article}
\usepackage{authblk}
\usepackage[utf8]{inputenc}
\usepackage{amsmath,amssymb,amsfonts,amsthm}
\usepackage{physics}
\usepackage{dsfont}
\usepackage{graphicx}
\usepackage{xcolor}
\usepackage{cancel}
\usepackage[numbers,sort&compress,square]{natbib}
\usepackage{color}
\usepackage{bbm}
\usepackage[linktocpage]{hyperref}
\usepackage{appendix}
\hypersetup{colorlinks=true,citecolor=blue,linkcolor=blue, urlcolor=blue, breaklinks=true}
\usepackage[eulergreek]{sansmath}
\usepackage{mathtools}
\usepackage{scalerel}

\usepackage{ulem}

\usepackage{bm} 
\usepackage{subfigure} 
\usepackage{environ}
\usepackage{url}
\usepackage{hyperref}
\usepackage[margin=1in]{geometry}

\usepackage{environ}
\NewEnviron{eqs}{%
\begin{equation}\begin{split}
    \BODY
\end{split}\end{equation}}

\newcommand{\Z}{{\mathbb Z}}

\newcommand{\bk}{{\bf k}}

\graphicspath{{./Figures/}}

\usepackage{adjustbox}

\begin{document}

\title{Remarks on invertible phases with non-onsite symmetry}
\author[1, *]{Ryohei Kobayashi}
\author[2,3, $\dagger$]{Kansei Inamura}
\author[4, $\ddag$]{Ken Shiozaki}

\affil[1]{Department of Applied Physics, The University of Tokyo, Tokyo 113-8656, Japan}
\affil[2]{Mathematical Institute, University of Oxford,
Andrew Wiles Building, Woodstock Road, Oxford, OX2 6GG, UK}
\affil[3]{Rudolf Peierls Centre for Theoretical Physics, University of Oxford,
Parks Road, Oxford, OX1 3PU, UK}
\affil[4]{Center for Gravitational Physics and Quantum Information,
Yukawa Institute for Theoretical Physics, Kyoto University, Kyoto 606-8502, Japan}

\renewcommand*{\thefootnote}{*}
\footnotetext[1]{Contact author: ryohei.k@ap.t.u-tokyo.ac.jp}
\renewcommand*{\thefootnote}{$\dagger$}
\footnotetext[1]{Contact author: kansei.inamura@physics.ox.ac.uk}
\renewcommand*{\thefootnote}{$\ddag$}
\footnotetext[1]{Contact author: ken.shiozaki@yukawa.kyoto-u.ac.jp}

\renewcommand{\thefootnote}{\arabic{footnote}}

\date{\today}

\maketitle

\begin{abstract}
We study invertible phases protected by non-onsite symmetries carrying nontrivial lattice anomaly indices. While lattice anomalies are usually viewed as obstructions to symmetric short-range-entangled (SRE) states, we show that anomalous lattice symmetries can nevertheless admit invertible phases that are not SRE, and can moreover shift the set of symmetry-compatible invertible phases.
In particular, we consider a fermionic $\mathbb Z_4^F$ symmetry in (2+1) dimensions. For an onsite $\mathbb Z_4^F$ symmetry, symmetric invertible phases have integer chiral central charge $c_-\in\mathbb Z$, whereas a non-onsite symmetry with nontrivial lattice anomaly index obstructs all such phases despite having trivial continuum 't Hooft anomaly. We resolve this by constructing an exact exponentially quasi-local $\mathbb Z_4^F$ symmetry of a $p+ip$ superconductor with $c_-=1/2$. We compute its lattice anomaly, and find that the symmetry indeed forbids $c_-\in \mathbb{Z}$ invertible states. 
The allowed invertible phases are consequently shifted to $c_-\in\mathbb Z+1/2$.
By gauging the fermion parity of the $p+ip$ superconductor, one obtains an Ising topological order enriched by a $\mathbb Z_4$ symmetry. We describe the corresponding symmetry structure in terms of a fusion 2-category.
We further show that the $p+ip$ state admits an enlarged non-onsite $U(1)^f$ symmetry, providing an analogue of fractional quantum Hall response in an invertible phase. This non-onsite $U(1)^f$ symmetry again forbids Chern insulators carrying $c_-\in\mathbb{Z}$, while being compatible with invertible states with $c_-\in\mathbb{Z}+1/2$. This result motivates a systematic study of new classes of invertible phases and spin liquids enriched by non-onsite symmetries.

\end{abstract}

\tableofcontents

\section{Introduction}

Invertible phases provide one of the simplest settings in which symmetry and topology can be studied in interacting many-body systems. An invertible phase is a gapped phase that admits an inverse under stacking, and stable equivalence classes of invertible phases generally form a discrete Abelian group. In the presence of an onsite symmetry, short-range-entangled (SRE) states give rise to symmetry-protected topological (SPT) phases~\cite{chen2013SPT}. More generally, invertible phases also include intrinsically chiral states, which may not be SRE even in the absence of symmetry.

In parallel with the classification of invertible and SPT phases with symmetry, there has recently been substantial progress toward understanding the dynamical consequence of symmetries, namely the characterization of anomalies, directly at the lattice systems~\cite{else2014,seiberg2022lsm,Seifnashri2024lsm,kawagoe2025anomaly,kapustin2025highersymmetries,seifnashri2025disentangling,shirley2025QCA,Feng2026higherformanomalies,Feng2026onsite,Tu2026anomalies,Kapustin2025LSM,Kobayashi2026generalizedstatistics,czajka2025anomalieslatticehomotopyquantum}. 
In this approach, an anomaly is characterized by indices valued in group cohomology, defined microscopically from given exact lattice symmetry operator. The lattice anomalies also serve as obstruction to onsite realization of the symmetry, therefore the anomalous symmetries are necessarily non-onsite~\cite{shirley2025QCA, Tu2026anomalies, kapustin2025highersymmetries}.
Fermionic lattice systems exhibit an especially rich structure of anomalies, see e.g.,~\cite{Chatterjee2025quantum,Pace2025Tduality,Seiberg2025LSMCPT,seiberg2026torikleinbottlesmodulo,fidkowski2025noninvertiblebosonicchiralsymmetry,thorngren2026chirallatticegaugetheories,Seiberg2024Majorana,Gioia2026exact,berkowitz2024exactlatticechiralsymmetry,Kim2026fermi,Pace2025parity,liu2026anomaliesquantumspinsystems,lu2026fermionicvillainmodelexact,lewsmith2026infiniteorderlatticechiralanomalies,dharanikota202611dlatticediracfermions, chavda2026fermionicanomalies} for recent discussions of fermionic anomalies in the lattice models.

Invertible phases are conventionally classified in the presence of an onsite symmetry. An onsite symmetry always admits a symmetric product state that provides a canonical choice of the trivial gapped phase, and each invertible phase can therefore be canonically labeled by an element of the classification group.

The situation becomes subtler when the symmetry itself is non-onsite. One can still classify invertible phases under symmetric finite-depth circuits, in the stable setting where we allow the stacking of ancillary product states carrying an onsite action of the symmetry. Suppose first that the non-onsite symmetry is \textit{onsiteable}, namely, that after adding trivial ancillas it can be disentangled into an onsite symmetry by conjugation with a finite-depth circuit~\cite{seifnashri2025disentangling, Feng2026onsite}. In that case, the resulting classification of invertible phases is naturally relative rather than absolute. The reason is that the original non-onsite symmetry need not preserve any product state, so there is no canonical choice of a trivial phase. One may choose a disentangling unitary that transforms the symmetry into an onsite form, so that one obtains an absolute classification relative to the corresponding trivial state. However, such a disentangling unitary is generally not unique. In particular, when there is a nontrivial SPT phase in the classification, one can redefine a disentangling unitary by multiplying an SPT entangler, therefore different choices can shift the identification of the trivial phase. Thus, without specifying a particular transformation into an onsite form, the classification should be regarded as relative.

When the non-onsite symmetry is not onsiteable, the situation can be considerably more exotic. In particular, a nontrivial lattice anomaly can obstruct every symmetric SRE state. Nevertheless, as we demonstrate in this work, such a symmetry may still preserve an invertible phase that is not SRE. There is then no symmetric SRE reference state from which the classification can be reduced to the case with onsite symmetry. This raises a natural question: what kinds of invertible phases can be compatible with a non-onsite symmetry?

In this work, we show that anomalies of lattice symmetries can do more than simply forbid SRE states: it can shift the set of allowed invertible phases. We identify a class of invertible phases that exists intrinsically in the presence of non-onsite symmetry with nontrivial anomaly indices, and has no counterpart for an onsite realization of the same symmetry. An interesting example occurs for fermionic $\Z_4^F$ symmetry in (2+1) dimensions.

For an onsite $\Z_4^F$ symmetry, symmetric invertible phases are classified by an integer chiral central charge, $c_- \in \Z$, so that the $p+ip$ superconductor with $c_-=1/2$ is incompatible with an onsite $\Z_4^F$ symmetry (see Appendix \ref{app:anomaly}).
On the other hand, a non-onsite $\Z_4^F$ symmetry can carry a nontrivial lattice anomaly index~\cite{chavda2026fermionicanomalies}; this anomaly index forbids any symmetric SRE state and, moreover, obstructs invertible states with integer chiral central charge $c_-\in\Z$.
Curiously, the corresponding continuum 't Hooft anomaly of $\Z_4^F$ is nevertheless trivial~\cite{Garcia_Etxebarria_2019}. At first sight, these observations might suggest that such a lattice symmetry cannot preserve any invertible phase at all.

We resolve this puzzle by showing that the non-onsite $\Z_4^F$ symmetry with lattice anomaly index instead admits invertible phases with $c_- \in \Z+1/2$.
In particular, we explicitly construct an exact non-onsite $\Z_4^F$ symmetry of a $p+ip$ superconductor with $c_-=1/2$. The symmetry is exponentially quasi-local and satisfies the $\Z_4^F$ algebra exactly. We further compute its lattice anomaly index directly, and demonstrate that it is nontrivial. That symmetry excludes the conventional $c_-\in\Z$ invertible phases, and in fact selects a shifted family with half-odd-integer chiral central charge. By stacking with ordinary onsite $\Z_4^F$ invertible phases, one obtains the full shifted set $c_-\in\Z+1/2$.

This phenomenon is qualitatively different from the ambiguity associated with an onsiteable non-onsite symmetry. In particular, unlike the onsitebale symmetry, the symmetric invertible states cannot be transformed into the known one by action of any finite-depth circuit or quantum cellular automata (QCA) acting within finite-dimensional onsite Hilbert space.\footnote{A QCA disentangling a $p+ip$ superconductor with an infinite-dimensional onsite Hilbert space has recently been constructed in \cite{jones2026QCA}.}

We then study the symmetry enrichment of the Ising topological order obtained by gauging fermion parity in the $p+ip$ superconductor with $\Z_4^F$ symmetry. The resulting Ising topological order possesses a $\Z_4$ symmetry. As a consequence of the lattice anomaly, the quotient symmetry $\Z_2=\Z_4^F/\Z_2^F$ is extended by a condensation defect associated with the emergent fermion. This condensation defect generates a faithful $\Z_2$ symmetry at the microscopic lattice level, although its realization becomes non-faithful in the IR Ising TQFT.
This microscopic UV $\Z_4$ symmetry induces a nontrivial symmetry structure in the IR Ising TQFT. We describe this UV-induced symmetry structure in terms of an example of fusion 2-categories constructed in \cite{douglas2018fusion2categories}, which is defined from the $\Z_4$-crossed braided Ising category.

We further find that the $p+ip$ superconductor possesses an enlarged exact non-onsite $U(1)^f$ symmetry, of which the $\Z_4^F$ symmetry discussed above is a subgroup. This again leads to behavior that is impossible for an onsite $U(1)^f$ symmetry. For an onsite charge-conserving $U(1)^f$ symmetry, invertible phases are integer quantum Hall states realized in Chern insulators. In contrast, the non-onsite $U(1)^f$ symmetry of the $p+ip$ state forbids the Chern insulator states and exhibits more exotic symmetry fractionalization pattern: the projective action of symmetry operators truncated to a disk acts as a Majorana translation along its boundary. This suggests that a $\pi$ vortex for $U(1)$ symmetry, at which such a Majorana translation terminates, binds a Majorana zero mode and carries topological spin $h_v=1/16$.
Since the spin of a $\pi$ flux is related to the Hall response, this points to an effective half-quantized response, $\sigma_{xy}=1/2,$
and suggests an analogue of the fractional quantum Hall effect within an invertible phase.

In the current paper, we do not establish here a microscopic characterization of fractional Hall transport for the lattice model. We rather demonstrate the corresponding half-quantized response at the level of continuum quantum field theory, using $U(1)^f$ symmetry fractionalization in a fermionic topological field theory coupled to a twisted spin structure~\cite{kobayashi2026sixteenfoldway}. In this description, the $\pi$ flux of the $U(1)^f$ symmetry carries topological spin $1/16$, and coupling the $U(1)$ background to the corresponding $\Z_2$ one-form symmetry produces a Chern-Simons response with $\sigma_{xy}=1/2. $
Thus, while a microscopic transport interpretation remains to be developed, the continuum theory realizes the fractional response implied by the non-onsite lattice $U(1)^f$ symmetry.

Our results therefore suggest a broader perspective on anomalies on the lattice. Rather than serving only as obstructions to symmetric SRE, anomalies can modify the set of invertible phases compatible with symmetry. In the examples studied here, the anomaly shifts the allowed chiral central charge. It would be interesting to understand more generally how microscopic anomaly indices determine such anomaly-shifted classifications of invertible phases.

The remainder of this paper is organized as follows. In Sec.~\ref{sec:Z4F in 2+1D}, we construct the exact non-onsite $\Z_4^F$ symmetry of the (2+1)D $p+ip$ superconductor, introduce momentum-space invariants of the symmetry operator, and compute its lattice anomaly through the boundary fermionic QCA obtained by truncating the symmetry to a half-space. In Sec.~\ref{sec:Ising}, we describe the $\Z_4$-enriched Ising topological order obtained by gauging the fermion parity in $p+ip$ superconductor.
In Sec.~\ref{sec:U1f}, we extend the symmetry to $U(1)^f$ and provide a continuum description of its fractional Hall response. Reviews of the relevant background material and technical details are relegated to the appendices.

\section{$\Z_4^F$ symmetry of $p+ip$ superconductor}
\label{sec:Z4F in 2+1D}

\subsection{$p+ip$ superconductor with exact $\Z_4^F$ symmetry}
We begin by constructing an exact $\Z_4^F$ symmetry of the standard $p+ip$ superconductor. Consider a gapped BdG Hamiltonian in (2+1) dimensions,
\begin{align}
\widehat{H}
=
\frac{1}{2}\sum_{\bf k}
\Psi_{\bf k}^{\dagger}
\mathcal{H}_{\bf k}
\Psi_{\bf k},
\qquad
\Psi_{\bf k}
=
\begin{pmatrix}
c_{\bf k}\\
c^\dagger_{-\bf k}
\end{pmatrix}~,
\end{align}
with the BdG redundancy $\tau_x \mathcal{H}_{\bf k}^{*}\tau_x= -\mathcal{H}_{-\bf k}.$
We take
\begin{align}
\mathcal{H}_{\bf k}
=
(\sin k_x)\tau_x
-
(\sin k_y)\tau_y
+
(m-\cos k_x-\cos k_y)\tau_z~,
\label{eq:p+ip}
\end{align}
with $\{\tau_x, \tau_y, \tau_z\}$ the Pauli matrices.
For $0<m<2$, this Hamiltonian has BdG Chern number $\operatorname{Ch}(\mathcal{H}_{\bf k})=1$, and describes the $p+ip$ superconductor with chiral central charge $c_-=1/2$.

Let
\begin{align}
Q_{\bf k}
=
\frac{\mathcal{H}_{\bf k}}
{\sqrt{\mathcal{H}_{\bf k}^{2}}}
\end{align}
be the spectrally flattened Hamiltonian, so that $Q_{\bf k}^{2}=1$. We define a symmetry operator $\widehat{U}$ by the action on the Nambu spinor {$\widehat U \Psi^\dagger_{\bf k} \widehat U^\dagger = \Psi^\dagger_{\bf k} U_{\bf k}$ with }
\begin{align}
U_{\bf k}
=
-iQ_{\bf k}~.
\label{eq:def of Z4F symmetry of p+ip}
\end{align}
Since $Q_{\bf k}$ commutes with $\mathcal{H}_{\bf k}$, this transformation leaves the Hamiltonian invariant,
\begin{align}
U_{\bf k}\mathcal{H}_{\bf k}U_{\bf k}^{\dagger}
=
\mathcal{H}_{\bf k}~.
\end{align}
Moreover,
\begin{align}
U_{\bf k}^{2}=-1~.
\end{align}
The BdG redundancy implies $\tau_xQ_{\bf k}^{*}\tau_x=-Q_{-\bf k},$ and therefore
\begin{align}
U_{-\bf k}
=
\tau_xU_{\bf k}^{*}\tau_x~,
\end{align}
{so that this action preserves the Nambu redundancy $\Psi^\dagger_{\bf k}=\Psi_{-{\bf k}}^T\tau_x$ and hence defines a consistent transformation of the physical fermion operators.}
Due to $U_{\bk}^2=-1$, 
the symmetry operator generates a $\Z_4^F$ symmetry
\begin{align}
\widehat U^{2}=(-1)^F~.
\end{align}

Although the symmetry is not onsite, it remains exponentially quasi-local. Indeed, $Q$ is obtained by spectral flattening of a local gapped Hamiltonian. Its real-space matrix elements therefore decay exponentially with distance,
\begin{align}
|Q_{{\bf r},{\bf r}'}|
\lesssim
e^{-|{\bf r}-{\bf r}'|/\xi}~,
\end{align}
for some correlation length $\xi$. Consequently, the action of $\widehat U$ on a local fermion operator has exponentially decaying spatial tails.

\subsection{Anomaly index of $\Z_4^F$ symmetry}
Here we compute the lattice anomaly index of the $\Z_4^F$ symmetry introduced in \eqref{eq:def of Z4F symmetry of p+ip}.
Let us consider a fermionic lattice system with internal $G_f$ symmetry, characterized by a central extension $\Z_2^F\to G_f\to G_b$.
For a given $G_f$ symmetry $U(g)$ with $g\in G_b$ generated by a finite-depth circuit, one defines an anomaly index by truncating the symmetry operator onto a disk $A$ to define $U_A(g)$, and evaluating its projective action
\begin{align}
    U_A(g)U_A(h) = \Omega_{\partial A}(g,h)[(-1)^{F_A}]^{\omega_2(g,h)}U_A(gh)~,
\end{align}
where $[\omega_2]\in H^2(BG_b,\Z_2)$ is the extension class that characterizes the symmetry $\Z_2^F\to G_f\to G_b$.
The anomaly index $[n_2]\in H^2(BG_b,\Z_2)$ is defined by the fermionic QCA index~\cite{Fidkowski2019fermionicQCA} of the 1D unitary $\Omega_{\partial A}(g,h)$ localized along the boundary~\cite{chavda2026fermionicanomalies},
\begin{align}
    n_2(g,h) = \begin{cases}
0, &  \text{$\text{Ind}_F(\Omega_{\partial A}(g,h)) \in \mathbb{Q}_+$}~,\\
1, & \text{$\text{Ind}_F(\Omega_{\partial A}(g,h)) \in \sqrt{2}\mathbb{Q}_+$}~.
\label{eq:fQCAindex}
\end{cases}
\end{align}
See Appendix \ref{app:anomaly} for details.
The nontrivial index $[n_2]$ forbids SRE states, and in the case of $\Z_4^F$ symmetry, the nontrivial $[n_2]=[\omega_2]$ further forbids an invertible state with $c_-\in\Z$. Fermionic anomalies are generally labeled by a layer of cohomology indices, and this $[n_2]$ index is dubbed a Majorana layer~\cite{Kapustin:2017jrc, thorngren2019anomaliesbosonization,Delmastro2021global}.
Below we evaluate this anomaly index of our $\Z_4^F$ symmetry.

To compute this anomaly index for the quasi-local operator $\widehat{U}$ defined by \eqref{eq:def of Z4F symmetry of p+ip}, one first needs to take a truncation of the symmetry operator. Let $\Pi_A$ be the one-particle projector onto a disk $A$ in the 2D space. Then define
$\widehat{Q}_A:= \Pi_A \widehat{Q} \Pi_A$, where $\widehat{Q}$ corresponds to the flattened Hamiltonian: 
\begin{align}
    \widehat{Q} =\frac{1}{2}\sum_{\bk} \Psi_{\bk}^\dagger Q_{\bk} \Psi_{\bk}~, \quad Q_{\bk}=\frac{\mathcal{H}_{\bk}}{\sqrt{\mathcal{H}_{\bk}^2}}~.
\end{align}
We define the truncated symmetry by
\begin{align}
    \widehat{U}_A = \exp\left(-\frac{\pi i}{2}\widehat{Q}_A\right)~.
\end{align}
Due to the exponential locality of $\widehat{Q}$, the unitary $\widehat{U}_A$ naturally becomes a truncation within the region $A$ in the sense that
\begin{align}
    ||\widehat{U}_A\mathcal{O}\widehat{U}^\dagger_A - \widehat{U}\mathcal{O}\widehat{U}^\dagger|| \lesssim e^{-\frac{r}{\xi}}
\end{align}
for a local operator $\mathcal{O}$ supported a distance $r$ inside the disk $A$.

It is convenient to introduce a projection operator $\hat{P}$ onto the lower-energy band
and its truncation by
\begin{align}
    \widehat{P}=\frac{1-\widehat{Q}}{2}~, \quad \widehat{P}_A
 = \Pi_A \widehat{P}_A \Pi_A~,
 \end{align}
which is a projection onto flattened lower-energy BdG bands.
Then, the anomaly index corresponds to the quasi-1D operator $\Omega_{\partial A}$,
\begin{align}
    \Omega_{\partial A} = \widehat{U}_A^2 (-1)^{F_A} = e^{2\pi i \widehat{P}_A}~.
\end{align}
Since the operator $\widehat{P}$ is a projector, its eigenvalue of the truncation $\widehat{P}_A$ on the states localized deep in the bulk of $A$ is accumulated at either 0 or 1. This ensures that $e^{2\pi i \widehat{P}_A}$ is exponentially localized at the boundary.
Let us now choose the region as
\begin{align}
    A = \{x\ge 0\}~,
\end{align}
then the operator $\widehat{P}_A$ has translation invariance along the $y$ direction. Using the momentum space representation ${P}_{\bk}$, we define
\begin{align}
    P_{x, x'}(k_y=k) = \int^{\pi}_{-\pi} \frac{dk_x}{2\pi} e^{ik_x(x-x')} {P}_{k_x,k}~.
\end{align}
One can then define its half-space truncation and the boundary unitary by
\begin{align}
    \widehat{P}(k) = \Pi_A P(k_y=k) \Pi_A~, \quad v(k) = e^{2\pi i \widehat{P}(k)}~.
\end{align}
Let us introduce the winding number of $v(k)$ by
\begin{align}
    \chi_{\partial A} = \frac{1}{2\pi i}\int^{\pi}_{-\pi} dk \mathrm{Tr}_{x\ge 0}[v(k)^{-1}\partial_kv(k)]~.
\end{align}
The one-dimensional winding invariant $\chi_{\partial A}$ essentially counts the number of edge modes along $\partial A$. Although the half-space truncated operator $\widehat P(k_y)=\Pi_A P(k_y)\Pi_A$ is no longer a projector, it remains Hermitian and satisfies
\begin{align}
0\leq \widehat P(k_y)\leq 1~.
\end{align}
Its bulk spectrum accumulates at $0$ and $1$, inherited from the eigenvalues of the bulk projector, whereas eigenvalues lying strictly inside the interval $(0,1)$ correspond to modes localized near the boundary $\partial A$.

As $k_y$ is varied across the Brillouin zone, these boundary eigenvalues can exhibit a spectral flow between the two bulk spectral sectors at $0$ and $1$. The winding invariant $\chi_{\partial A}$ counts the net spectral flow of edge modes, namely the number of boundary eigenvalues flowing from $0$ to $1$ minus those flowing from $1$ to $0$. By the bulk-boundary correspondence, its net spectral flow is equal to the Chern number of the bulk occupied-band projector,
\begin{align}
\chi_{\partial A} = \text{Ch}(\widehat P)~,
\end{align}
up to the choice of orientation of $\partial A$. Our $p+ip$ Hamiltonian therefore has $\chi_{\partial A}= 1$.

It is known that the winding number associated with the translationally invariant quantum walk characterizes the 1D QCA index \cite{GNVW2012, Zimbors2022}. When the quantum walk is expressed by the BdG representation of free fermions, the relation between the winding number and the fermionic QCA index is given by
\begin{align}
    [\text{Ind}_F(v(k))]^2 = 2^{\chi_{\partial A}}~,
\end{align}
where the square accounts for the doubled degrees of freedom in the BdG Hamiltonian.\footnote{
Indeed, consider the Majorana translation of a Majorana chain with two Majorana fermions per site $\{a_j,b_j\}$. The unitary in the Fourier transformed Majorana basis $\{a_k, b_k\}$ is implemented by a matrix
\begin{align}
    v(k) = \begin{pmatrix}
        \ & 1 \\
        e^{ik} & \
    \end{pmatrix}~,
\end{align}
then $\det(v(k)) = -e^{ik}$, which leads to the winding number 1. This corresponds to the fermionic QCA with the index $\sqrt{2}$. }
This implies that the anomaly index becomes nontrivial $n_2(1,1)=1$, which implies $[n_2]=[\omega_2]$ is nontrivial. Therefore the $\Z_4^F$ symmetry forbids an invertible state with $c_-\in\Z$, but nevertheless admits one with $c_-\in \Z+1/2$.
In Appendix \ref{app:Z4 in 1+1D}, we also discuss much simpler example of similar phenomena in (1+1)D where the anomaly index of a symmetry forbids SRE, while admitting the Kitaev's Majorana chain.

\section{$\Z_4$ symmetry of Ising topological order}
\label{sec:Ising}
Once we gauge fermion parity in the $p+ip$ superconductor with $\Z_4^F$ symmetry, we obtain the Ising topological order enriched by some symmetry.
In this section, we discuss the algebraic structure of this symmetry and its action on the Ising TQFT in the IR.

\subsection{UV symmetry}
\label{sec: UV symmetry}
We first consider the UV microscopic symmetry of the gauged model on the lattice.
Since we have gauged the fermion parity symmetry, the resulting model has a $\Z_2$ 1-form symmetry generated by a fermion line $\psi$.
Consequently, this model also has a 0-form symmetry generated by the condensation defect $\mathcal{C}_{\psi}$ of the fermion~\cite{Roumpedakis2023}.
If we enforce the local Gauss law upon gauging $\Z_2^F$, the fermion line $\psi$ becomes topological and its condensation defect generates a $\Z_2$ symmetry.
On the other hand, if we do not impose the local Gauss law constraint, the symmetry may become larger.
For instance, in the toric code model, the condensation defect corresponds to the em-exchange symmetry that may be implemented by a $\Z_4$ symmetry operator on a tensor product Hilbert space~\cite{Tu2026anomalies, shirley2025QCA, kobayashi2026exactem}.
In what follows, we will focus on the case where the local Gauss law constraint is imposed at the level of the Hilbert space.
In this case, the symmetry generated by the condensation defect is $\Z_2$.

In addition to the above symmetries, the gauged model also has a residual 0-form symmetry of the original $\Z_4^F$ symmetry.
Naively, one may expect that this residual symmetry is described by $\Z_2 = \Z_4^F / \Z_2^F$ because we have gauged the $\Z_2^F$ subgroup of $\Z_4^F$.
However, we claim that this is not the case.
More specifically, we claim that the residual symmetry is extended non-trivially by the $\Z_2$ symmetry generated by the condensation defect.
As a consequence, the total 0-form symmetry group becomes $\Z_4$.

Indeed, when the anomaly index of the Majorana layer is present $[n_2]\in H^2(BG_b,\Z_2)$ in the original (2+1)D invertible phase with $G_f$ symmetry, fusing in parallel a pair of $g,h\in G_b = G_f/\Z_2^F$ symmetry operators in the invertible phase gives the Kitaev's Majorana chain supported at the defect surface~\cite{WangGu2020}. After gauging $\Z^F_2$ symmetry, the Kitaev's Majorana chain is realized as a condensation defect of the fermion $\psi$~\cite{Barkeshli2023codim2, Barkeshli2024higherfermion}. Therefore the 0-form $G_b$ symmetry of the resulting topological order is extended by a condensation defect $\mathcal{C}_{\psi}$ according to the extension class $[n_2]\in H^2(BG_b,\Z_2)$; this phenomenon has been pointed out in~\cite{Bulmash2022cascade}.

Summarizing, the 0-form symmetry defects are generated by $\mathcal{D}_g$ with $g$ a generator of $\Z_4$, and it fuses as
\begin{equation}
\mathcal{D}^2_g = \mathcal{C}_\psi.
\end{equation}
The condensation defect $\mathcal{C}_\psi$ is necessarily terminated at a non-invertible line $\sigma_{\mathrm{UV}}$, $\text{Hom}(\mathcal{C}_{\psi}, 1)=\{\sigma_{\mathrm{UV}} \}$.
Equivalently, a trivalent junction of two non-trivial surfaces $\mathcal{D}_g$ and one trivial surface $1$ must support a non-invertible line $\sigma_{\mathrm{UV}}$.
This non-invertible junction braids non-trivially with the fermion line $\psi$, reflecting the fact that the original symmetry was a non-trivial extension of $\Z_2$ by $\Z_2^F$.

The non-trivial braiding between a fermion line $\psi$ and a junction $\sigma_{\mathrm{UV}}$ is reminiscent of the symmetry fractionalization.
We will discuss more details on this point in later subsections.

\paragraph{Categorical description}
The above algebraic structure of the UV symmetry corresponds to an example of a fusion 2-category discussed in \cite{douglas2018fusion2categories}, obtained by (semisimple completion of) the following $\Z_4$-crossed braided Ising category,
\begin{align}
 C_0 = \{1,\psi\}~, \quad  C_1 = 0~, \quad C_2 = \{\sigma_{\mathrm{UV}}\}~, \quad C_3 = 0~,
\end{align}
where $C_j$ corresponds to the termination of $j\in \Z_4$ 0-form symmetry defects; $C_j = \text{Hom}(\mathcal{D}_j, 1)$ with $\mathcal{D}_1=\mathcal{D}_g, 
\mathcal{D}_2=\mathcal{C}_\psi$. The sector $C_1, C_3$ are empty, implying that this $\Z_4$-crossed category is non-faithfully graded. This physically corresponds to the fact that the termination of the symmetry operator $\mathcal{D}_g$ is not topological. Since the defect $\mathcal{D}_2=\mathcal{C}_{\psi}$ is a non-faithful symmetry operator in the TQFT, the endpoint operator $\sigma_{\mathrm{UV}}$ is still interpreted as a genuine line operator $\sigma$ in the IR; see Sec.~\ref{sec: IR symmetry} for more details. This reflects that the Ising TQFT is self-dual under 1-gauging the $\Z_2$ 1-form symmetry of the fermion $\psi$, and the non-invertible topological line $\sigma$ is obtained by half-higher-gauging the fermion $\psi$~\cite{choi2023noninvertiblegausslawaxions}.

Precisely speaking, the UV symmetry operators on the lattice may not form a fusion 2-category on the nose.
This is because when the lattice anomaly is present, the junction of symmetry operators on the lattice may involve lattice translations, which go beyond a fusion 2-categorical description.
Nevertheless, in the IR, the symmetry operators induced by the microscopic UV symmetry under the RG flow should be described by a fusion 2-category.
For brevity, we refer to this induced symmetry structure as the \textit{UV-induced symmetry} of the Ising TQFT.
We claim that this UV-induced symmetry structure is described by the semisimple completion of the $\mathbb{Z}_4$-crossed braided Ising category introduced above.
Notably, this fusion 2-category is not monoidally equivalent to the 2-category of twisted 2-group-graded 2-vector spaces~\cite{douglas2018fusion2categories}, which means that the symmetry structure is not a 2-group, even though both the 0-form and 1-form symmetries are invertible.

We emphasize that the UV-induced symmetry should not be identified with the intrinsic symmetry structure of the IR Ising TQFT, which is described by the fusion 2-category $\mathrm{Mod}(\mathrm{Ising})$. Rather, the UV-induced symmetry structure of the IR theory retains information about the microscopic UV symmetry, including its non-faithful realization under RG flow.
The relation between the UV-induced symmetry and the IR intrinsic symmetry of the Ising TQFT will be discussed in the next subsection.

\subsection{IR symmetry}
\label{sec: IR symmetry}
We now discuss the intrinsic symmetry of the IR Ising TQFT, which we refer to as the IR symmetry.
We will also describe the action of the UV-induced symmetry on the IR TQFT and briefly comment on a relation to the symmetry fractionalization.

The IR Ising TQFT has an intrinsic (non-invertible) 1-form symmetry described by the Ising modular tensor category.
This category has three simple objects $\{1, \psi, \sigma\}$ corresponding to three anyons.
The condensation completion of this category produces a fusion 2-category $\mathrm{Mod}(\mathrm{Ising})$~\cite{douglas2018fusion2categories}, which describes the entire IR symmetry.
This fusion 2-category has only one simple object, corresponding to the identity surface operator.
In particular, the Ising TQFT has no intrinsic 0-form symmetry.

Since the IR TQFT has no intrinsic 0-form symmetry, the UV 0-form symmetry $\Z_4$ must act trivially in the IR.
In other words, the UV symmetry defect $\mathcal{D}_g$ becomes a trivial defect in the IR.
On the other hand, the UV 1-form symmetry generated by the fermion line cannot become trivial in the IR because it is anomalous.
Hence, the fermion line in the UV must be identified with the one in the IR, both denoted by $\psi$.
Furthermore, since the condensation defect becomes trivial in the IR, the non-invertible termination $\sigma_{\mathrm{UV}}$ of the condensation defect in the UV flows to a genuine non-invertible line $\sigma$ in the IR.
Therefore, the UV-to-IR map of the symmetry category is given by
\begin{equation}
\mathcal{D}_g \mapsto 1, \qquad
\psi \mapsto \psi, \qquad
\sigma_{\mathrm{UV}} \mapsto \sigma.
\end{equation}
We note that a trivalent junction of two $\mathcal{D}_g$'s and one trivial surface is mapped to a non-invertible line $\sigma$.
On the other hand, a junction of two $\mathcal{D}_g$'s and one condensation surface $\mathcal{C}_{\psi}$ is mapped to either a trivial line or a fermion line.
The latter can always be made into a trivial line by attaching a fermion line to the junction in the UV.

Naively, one may expect that $\mathcal{D}_g$ generates a $\Z_2$ symmetry in the IR that fractionalizes a fermion $\psi$, since the junction $\sigma_{\mathrm{UV}}$ braids non-trivially with $\psi$.
However, this cannot be regarded as a symmetry fractionalization of a $\mathbb{Z}_2$ symmetry in the usual sense.
Indeed, recalling that the junction $\sigma_{\mathrm{UV}}$ originates from the endpoint of the condensation defect, one finds that the non-invertible line localized at the junction in the UV can be pulled off in the IR; that is, we have
\begin{equation}
\adjincludegraphics[valign=c, trim={10, 10, 10, 10}, scale=1]{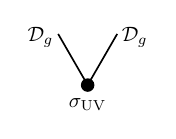}
=
\adjincludegraphics[valign=c, trim={10, 10, 10, 10}, scale=1]{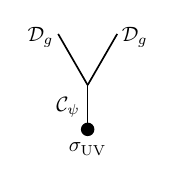}
\quad \xrightarrow{\text{UV to IR}} \quad
\adjincludegraphics[valign=c, trim={10, 10, 10, 10}, scale=1]{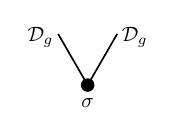}
=
\adjincludegraphics[valign=c, trim={10, 10, 10, 10}, scale=1]{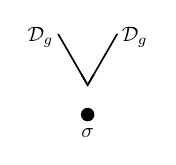}
\end{equation}
Here, we used the fact that the condensation defect $\mathcal{C}_{\psi}$ acts trivially in the IR.
The above equation implies that the symmetry defect admits a junction that braids trivially with $\psi$ in the IR TQFT, and hence there is no symmetry fractionalization in the usual sense.
This is consistent with the fact that the Ising TQFT cannot have a $\Z_2$ symmetry that fractionalizes $\psi$~~\cite{Barkeshli2019Symmetry}.
That said, one could still perhaps think of the above symmetry structure as an unconventional form of symmetry fractionalization.

\subsection{Relation to the classification of fermionic invertible phases}
Fermionic invertible phases with symmetry in 2+1 dimensions were classified in~\cite{Barkeshli:2021ypb} based on the classification of bosonic symmetry-enriched topological orders~\cite{Barkeshli2019Symmetry} obtained by gauging the fermion parity.
According to the classification in~\cite{Barkeshli:2021ypb}, fermionic invertible phases with $\Z_4^F$ symmetry must have integer chiral central charge $c_- \in \Z$.
The derivation of this result relies on the assumption that the bosonic topological order obtained by gauging $\Z_2^F \subset \Z_4^F$ has a $\Z_2$ 0-form symmetry that fractionalizes a fermion line $\psi$.
However, this assumption is valid only when the $\Z_4^F$ symmetry has no lattice anomaly.
Indeed, as discussed above, when the $\Z_4^F$ symmetry has a non-trivial lattice anomaly $[n_2] = [\omega_2]$, the 0-form symmetry after the gauging is described by $\Z_4$, and there is no $\Z_2$ symmetry fractionalization of $\psi$ in the usual sense.
Therefore, the classification in~\cite{Barkeshli:2021ypb} does not apply to the $\Z_4^F$ symmetry with a lattice anomaly.

When the $\Z_4^F$ symmetry has a non-trivial lattice anomaly, the symmetry after gauging the $\Z_2^F$ subgroup is described by (the semisimple completion of) a $\Z_4$-crossed braided Ising category discussed in Sec.~\ref{sec: UV symmetry}.
This symmetry is incompatible with the underlying topological order when $c_- \in \Z$.
Indeed, when $c_- \in \Z$, the condensation defect of $\psi$ acts as a non-trivial anyon permutation symmetry in the IR TQFT.
However, this anyon permutation symmetry does not have a square root.
Therefore, the $\Z_2$ symmetry generated by the condensation defect cannot be a $\Z_2$ subgroup of $\Z_4$.
This implies that the $\Z_4^F$ symmetry with a lattice anomaly only admits invertible phases with $c_- \in \Z + 1/2$.
This result aligns with a lattice analysis in Appendix~\ref{app:anomaly}.

\section{$U(1)^f$ symmetry of $p+ip$ superconductor}
\label{sec:U1f}

The $\Z_4^F$ symmetry discussed in the previous section is a subgroup of the $U(1)^f$ symmetry generated by
\begin{align}
    U^{\theta}_{\bk} = e^{-i\theta Q_{\bk}}~,
\end{align}
which obeys $\widehat{U}(\theta=\pi)=(-1)^F$, hence generates $U(1)^f$ symmetry of the $p+ip$ Hamiltonian \eqref{eq:p+ip}. 

For a given $U(1)^f$ symmetry operator, one can again define the index $[n_2]\in H^2(BU(1),\Z_2)$ by truncating the symmetry operator onto a disk region $A$ and
\begin{align}
    \widehat U_A(\theta)\widehat U_A(\theta') = \Omega_{\partial A}(\theta,\theta')[(-1)^{F_A}]^{\omega_2(\theta,\theta')}\widehat U_A(\theta+\theta')~,
    \label{eq:U1projective}
\end{align}
with $\theta,\theta'\in [0,\pi)$, and $\omega_2\in Z^2(BU(1),\Z_2)$ is an extension class characterizing $\Z_2^F\to U(1)^f\to U(1)^b$. Then the fQCA index of $\Omega_{\partial A}$ as in \eqref{eq:fQCAindex} again defines $[n_2]\in H^2(BU(1),\Z_2)$. This index is invariant under shifting the choices of the truncated symmetry operators along the boundary. The nontrivial class $[n_2]$ carried by the symmetry $\widehat U$ again forbids invertible states with $c_-\in\Z$; in particular, this non-onsite $U(1)$ symmetry does not preserve standard Chern insulators with onsite $U(1)$ symmetry carrying integer Hall response.

The boundary algebra \eqref{eq:U1projective} further suggests a physical interpretation of this lattice anomaly in terms of $U(1)$
vortices. For the nontrivial class $[n_2]=[\omega_2]$, the projective composition of the truncated $U(1)$ symmetry operators induces a Majorana translation along $\partial A$. The projective algebra suggests that such a boundary Majorana translation terminates at a $\pi$ symmetry vortex, and its endpoint is naturally expected to bind a Majorana zero mode. This motivates assigning the $\pi$ vortex the spin $h_v = 1/16$.
This is half the corresponding value $h_v=1/8$ for the $\pi$ flux of the integer quantum Hall state $U(1)_1$. Since the spin of the vortex is proportional to the Hall conductance, this observation suggests an effective fractional Hall response $\sigma_{xy} = 1/2$.

We note that this argument should not by itself be regarded as a microscopic characterization of a fractional quantum Hall response for the lattice model. In particular, we do not establish here a microscopic transport diagnostic or a many-body definition of Hall conductance directly from the non-onsite lattice $U(1)^f$ symmetry. Rather, the Majorana zero mode at the $\pi$ vortex of the $U(1)$ symmetry motivate the value $\sigma_{xy}=1/2$, which we substantiate below at the level of the continuum topological field theory by coupling the $U(1)^f$ background to the TQFT describing the $p+ip$ superconductor. Developing a fully microscopic diagnostic of this response, and clarifying in what sense an analogue of Hall transport can be formulated for such non-onsite symmetries on the lattice, would be interesting directions for future work.

\paragraph{Symmetry fractionalization of twisted spin TQFT}
We now provide a continuum description of the fractional Hall response $\sigma_{xy}=1/2$ for the $p+ip$ superconductor. This can be understood in terms of $U(1)^f$ symmetry fractionalization in a (2+1)D TQFT coupled to a twisted spin structure~\cite{kobayashi2026sixteenfoldway}. The TQFT of interest possesses a $\Z_2$ 1-form symmetry with 2-form background field $B$, and the obstruction to the twisted spin structure is given by $w_2+B$, where $w_2$ denotes the second Stiefel-Whitney class.

One can obtain the twisted spin TQFT in a following way~\cite{kobayashi2026sixteenfoldway}. Let us start with an Ising TQFT with anyons $\{1,\psi,\sigma\}$, and proliferate/condense the fermion $\psi$ along with inserting the $\sigma$ line along the Poincar\'e dual of $w_2$. The resulting theory has a topological line operator $\tau$ originating from $\sigma$ of the Ising; it is deconfined since the mutual braiding between $\sigma, \psi$ and the framing anomaly of $\psi$ cancels out. Although the line operator $\tau$ obeys the non-invertible fusion rule
$\tau\times\tau=1+f$
where $f$ is the \textit{local} fermion, the locality of $f$ implies that $\tau$ can be regarded as generating an anomalous $\mathbb{Z}_2$ 1-form symmetry~\cite{Bhardwaj2025fermionic}. 
Let us write its 2-form background as $B$. Since $\tau$ is inserted along the Poincar\'e dual of $w_2$, the theory is coupled to twisted spin structure with the obstruction $w_2+B$.

The $p+ip$ state on a torus is regarded as a superselection sector of this twisted spin TQFT that corresponds to fixed spin structure. One can couple $U(1)^f$ symmetry with the background gauge field $A$ to the twisted spin TQFT by setting ${dA} = \pi B$ mod $2\pi$, so that one naturally satisfies the Dirac quantization condition for the spin$^c$ structure: $dA=\pi w_2$.

The Hall conductance can then be computed through the response of the $\Z_2$ 1-form symmetry in the twisted spin TQFT.
The $\Z_2$ 1-form symmetry $\tau$ carries the 't Hooft anomaly described by a (3+1)D topological response, classified by the reduced bordism of 4-manifolds with twisted spin structure, which is $\Z_{16}$. The $\Z_2$ 1-form symmetry of our TQFT obtained from the Ising theory precisely generates this $\Z_{16}$ classification, which manifests itself as the spin 1/16 carried by $\tau$. The (3+1)D topological response is expressed as
\begin{align}
    \exp\left(\frac{2\pi i}{16} \int \hat B\cup \hat B\right)\times (-1)^{\text{Arf}(F_{\hat{B}})}~,
\end{align}
where $\hat B\in Z^2(M,\Z)$ is the integral lift of the background $B$, and $F_{\hat{B}}$ is its dual surface.

By plugging $dA=\pi B$ in this response, the resulting theta angle from the first term corresponds to the Chern-Simons response
\begin{align}
    \exp\left(\frac{1}{2} \frac{i}{4\pi} \int AdA\right)~,
\end{align}
with the Hall conductance $\sigma_{xy} = 1/2$.

\section{Discussions}

In this work, we have explored invertible phases protected by non-onsite symmetries carrying nontrivial lattice anomaly indices. Our main example is a non-onsite $\mathbb Z_4^F$ symmetry in (2+1) dimensions. We constructed an exact exponentially quasi-local realization of this symmetry in a $p+ip$ superconductor and showed that it carries the nontrivial Majorana-layer anomaly $[n_2]=[\omega_2]$. While this anomaly obstructs all symmetric invertible phases with $c_-\in\mathbb Z$, it is nevertheless compatible with the $p+ip$ state with $c_-\in \Z+1/2$. 

After gauging fermion parity, we found a $\Z_4$ symmetry of Ising topological order. The quotient $\mathbb Z_2=\Z_4^F/\Z_2^F$ symmetry is extended by the condensation defect $\mathcal C_\psi$, such that the symmetry defect satisfies $\mathcal D_g^2=\mathcal C_\psi$. Since $\mathcal C_\psi$ can terminate on the non-Abelian line $\sigma$, the junction of two $\Z_4$ symmetry defects can support a non-Abelian anyon. This provides an unconventional form of symmetry fractionalization that lies beyond the usual description in terms of Abelian anyons decorating junctions of symmetry defects. We described this structure by the semisimple completion of a non-faithfully graded $\mathbb Z_4$-crossed braided Ising category.

Our results raise several natural directions for future work. One particularly interesting direction is the systematic study of symmetry-enriched topological phases with non-onsite symmetry. Our Ising example demonstrates that non-onsite symmetry can allow symmetry junctions to support non-Abelian anyons, realizing an unconventional form of symmetry fractionalization. It would be interesting to develop a general categorical framework for such phases, in which non-Abelian anyons can be bound to symmetry junctions, and to understand systematically the underlying mechanism based on extensions of symmetry defects by condensation defects. Constructing microscopic lattice models of spin liquids that realize these structures would provide particularly concrete examples of topological orders intrinsically enriched by non-onsite symmetry.

A second direction concerns crystalline symmetries that are intrinsically non-onsite operations of lattice models. Our results suggest that allowing the microscopic crystalline symmetry action itself to carry nontrivial lattice anomaly data may enlarge the landscape of possible classification of crystalline phases and responses further. In particular, it would be interesting to ask whether lattice anomalies can shift the quantization of crystalline responses, or modify the possible quantum numbers or Majorana zero modes bound to dislocations or disclinations of the crystalline symmetry \cite{Manjunath2023nonperturbative, Zhang2023quantized, Zhang2023complete, Kobayashi2025FCI}. 

Finally, the enlarged non-onsite $U(1)^f$ symmetry of the $p+ip$ state suggests an unconventional response $\sigma_{xy}=1/2$ of an  invertible state. A fully microscopic formulation of this response remains to be developed. It would be particularly interesting to identify an appropriate lattice diagnostic of Hall transport for non-onsite continuous symmetries and to understand more generally how lattice anomaly data modifies the quantization of symmetry responses.

\section*{Acknowledgments}
We thank Takamasa Ando, Maissam Barkeshli and Po-Shen Hsin for discussions.
R.K. is supported by the Department of Applied Physics,
the University of Tokyo. K.I. is supported by
the Leverhulme-Peierls
Fellowship funded by the Leverhulme Trust and the EPSRC Open Fellowship EP/X01276X/1. 
KS is supported by JSPS KAKENHI Grant Nos.~JP22H05118,
JP26H01305, and JP26K00629.

\appendix

\section{Anomaly indices of lattice models}
\label{app:anomaly}

\subsection{Review of anomaly indices}

We first briefly review the lattice anomaly indices for fermionic symmetries introduced in \cite{chavda2026fermionicanomalies}, focusing on the Majorana-layer $[n_2]$ index that will be relevant in the main text. Consider a finite internal symmetry $G_f$ in (2+1)D, given by a central extension $\Z_2^F\to G_f\to G_b$, 
characterized by an extension class $[\omega_2]\in H^2(BG_b,\Z_2)$.
The symmetry operators labeled by $g,h\in G_b$ satisfy the group algebra
\begin{align}
\widehat{U}(g)\widehat{U}(h)
=
[(-1)^F]^{\omega_2(g,h)}
\widehat{U}(gh)~.
\label{eq:Gf algebra review}
\end{align}

To extract the lattice anomaly, we truncate the symmetry operator onto a disk region $A$, and denote the truncated operator by $\widehat{U}_A(g)$. Such a truncation exists since a fermionic QCA in two spatial dimensions is a finite depth circuit. \footnote{
In this paper, finite-depth circuits are not assumed to be strictly local; more generally, we allow the local gates to have exponentially decaying tails. Since the classification of one-dimensional (f)QCAs is robust under such exponentially decaying corrections \cite{ranard2026approximateqcasdimensionusing, Ranard2022converse}, we expect the corresponding anomaly indices to remain well defined and robust in this setting. This expectation is supported by the explicit $\mathbb{Z}_4^F$ example studied in the main text, for which the anomaly index can be computed directly despite the presence of exponential tails.
We emphasize that the absence of strict locality introduces additional subtleties in the precise definition of these anomaly indices, and we do not attempt a fully rigorous treatment of these issues in the present work. Accordingly, by an SRE state we also mean, more generally, a state that can be disentangled by a finite-depth quantum circuit whose local gates may have exponentially decaying tails, rather than by a strictly local circuit.
} 
The multiplication law of the truncated symmetry takes the form
\begin{align}
\widehat{U}_A(g)\widehat{U}_A(h)
=
\Omega_{\partial A}(g,h)
[(-1)^{F_A}]^{\omega_2(g,h)}
\widehat{U}_A(gh)~,
\label{eq:truncated symmetry review}
\end{align}
where $\Omega_{\partial A}(g,h)$ is a one-dimensional fermionic locality-preserving unitary, i.e., a fermionic quantum cellular automaton (fQCA), supported near the boundary $\partial A$.
A one-dimensional fQCA has an index valued in~\cite{Fidkowski2019fermionicQCA}
\begin{align}
\operatorname{Ind}_F(\Omega_{\partial A})
\in
\mathbb{Q}_{+}
\cup
\sqrt{2}\mathbb{Q}_{+}~.
\end{align}
The factor of $\sqrt{2}$ is intrinsically fermionic and is represented by translation of a single Majorana fermion. This allows us to define a $\Z_2$-valued 2-cochain
\begin{align}
    n_2(g,h) = \begin{cases}
0, &  \text{$\text{Ind}_F(\Omega_{\partial A}(g,h)) \in \mathbb{Q}_+$}~,\\
1, & \text{$\text{Ind}_F(\Omega_{\partial A}(g,h)) \in \sqrt{2}\mathbb{Q}_+$}~.
\label{eq:fQCAindexApp}
\end{cases}
\end{align}
Physically, $n_2(g,h)=1$ means that the projective multiplication of the two truncated symmetry operators leaves behind a Majorana translation along $\partial A$.
Associativity of the symmetry action implies $\delta n_2=0$. 
Moreover, changing the choice of truncation $\widehat{U}_A(g)$ can shift $n_2$ by a coboundary. Therefore, the invariant data associated with the symmetry operator is the cohomology class
\begin{align}
[n_2]\in H^2(BG_b,\Z_2)~.
\label{eq:n2 anomaly class}
\end{align}
We refer to this anomaly index as the Majorana layer. When $[n_2]$ is trivial, one can successively define further anomaly indices $[n_3]\in H^3(BG_b,\Z_2)$ and $[\nu_4]\in H^4(BG_b,\mathbb{R}/\mathbb{Z})$, but these higher layers will not be used in this paper.

\subsection{Obstruction to SRE}

Here we review why a nontrivial Majorana-layer index provides an obstruction to a symmetric SRE state. 
Suppose that a $G_f$-symmetric SRE state exists,
\begin{align}
\ket{\Psi}
=
V\ket{0}~,
\qquad
\widehat{U}(g)\ket{\Psi}
\propto
\ket{\Psi}~,
\label{eq:symmetric SRE assumption}
\end{align}
where $\ket{0}$ is a product state with the fermions in a Fock vacuum and $V$ is a finite-depth circuit. Conjugating the symmetry by $V$, we define $\widetilde{U}(g)
:=
V^\dagger
\widehat{U}(g)
V.$
The conjugated symmetry preserves the product state, $\widetilde{U}(g)\ket{0}
=
\ket{0}$,
while carrying the same anomaly class $[n_2]$, since conjugation by a finite-depth circuit does not change the fQCA index.

Let us truncate $\widetilde{U}(g)$ onto a disk $A$. Its action on the product state can create a one-dimensional invertible state near the boundary,
\begin{align}
\widetilde{U}_A(g)\ket{0}
=
\ket{0}
\otimes
\ket{\Phi_{g;\partial A}}~.
\label{eq:boundary state truncation}
\end{align}
A one-dimensional fermionic invertible phase without additional symmetry has a $\Z_2$ classification, generated by the Kitaev chain. We can therefore associate a 1-cochain
\begin{align}
\chi(g)\in\Z_2
\end{align}
to the boundary state $\ket{\Phi_{g;\partial A}}$, with $\chi(g)=0$ for the trivial phase and $\chi(g)=1$ for the Kitaev-chain phase.

Consider now the combination
\begin{align}
\widetilde{U}_A(g)
\widetilde{U}_A(h)
\widetilde{U}_A(gh)^\dagger
\ket{0}~.
\end{align}
From the boundary states in \eqref{eq:boundary state truncation}, the one-dimensional invertible phase produced by this action is labeled by
\begin{align}
\delta\chi(g,h)
=
\chi(g)+\chi(h)+\chi(gh)
\quad \mathrm{mod}\ 2~.
\end{align}
On the other hand, using the truncated symmetry algebra \eqref{eq:truncated symmetry review}, the same state is obtained by acting with $\Omega_{\partial A}(g,h)$ on the product state. A fermionic QCA carrying the $\sqrt{2}$ index transforms a trivial one-dimensional invertible state into the Kitaev-chain phase. Hence the resulting phase is labeled precisely by $n_2(g,h)$. We therefore obtain
\begin{align}
n_2
=
\delta\chi~.
\end{align}
It follows that $[n_2]=0$ for any symmetry-preserving SRE states.
Thus, a symmetry with $[n_2]\neq 0$ cannot preserve any SRE state.

\subsection{Dynamical consequence of $\Z_4^F$ symmetry}

We now specialize the above discussion to $G_f=\Z_4^F$ symmetry. 
The extension class $[\omega_2]
\in
H^2(B\Z_2,\Z_2)
=
\Z_2$ is nontrivial.
Writing $g$ for the generator of $G_b=\Z_2$, the $\Z_4^F$ symmetry algebra is
\begin{align}
\widehat{U}(g)^2
=
(-1)^F~.
\label{eq:Z4F algebra review}
\end{align}
We are interested in a non-onsite realization of this symmetry carrying the Majorana-layer anomaly $[n_2]
=
[\omega_2]
\neq
0$.
This anomaly index implies that such a $\Z_4^F$ symmetry cannot preserve an SRE state.

\subsubsection{Non-onsite $\Z_4^F$ symmetry: Invertible states with $c_-\in\Z+1/2$}

The Majorana layer anomaly index places a stronger constraint on the invertible phases compatible with the $\Z_4^F$ symmetry. We review that the non-onsite realization with $[n_2]=[\omega_2]$ instead excludes all symmetric invertible states with $c_-\in\Z$.

Suppose, for contradiction, that a symmetric invertible state $\ket{\Phi}$ with $c_-=k\in\Z$
were preserved by the non-onsite $\Z_4^F$ symmetry with $[n_2]=[\omega_2]$. We can stack $\ket{\Phi}$ with an ordinary onsite $\Z_4^F$-symmetric invertible state $\ket{\Psi_{-k}}$ carrying the opposite chiral central charge, $c_-=-k$.
For example, $\ket{\Psi_{-k}}$ can be constructed from ordinary charge-conserving Chern insulators, with $\Z_4^F$ realized as an onsite subgroup of $U(1)^f$.
The stacked state then has vanishing chiral central charge, $c_-=0$.
At the same time, stacking with an onsite symmetry does not modify the Majorana-layer anomaly, so the symmetry acting on the stacked system still satisfies $[n_2]=[\omega_2]\neq 0$.
Since the stacked state is an invertible phase with vanishing chiral central charge, it is an SRE. This contradicts the obstruction to SRE from $[n_2]$. We therefore conclude that
\begin{align}
c_-\notin\Z
\label{eq:integer c obstruction}
\end{align}
for an invertible state preserving the non-onsite $\Z_4^F$ symmetry with $n_2=\omega_2$.

This does not mean that the symmetry forbids all invertible states. As demonstrated by the main text, the $p+ip$ superconductor with $c_-=1/2$
admits an exact exponentially quasi-local $\Z_4^F$ symmetry carrying precisely the nontrivial Majorana-layer index $[n_2]=[\omega_2]$.
Thus, the anomalous symmetry is compatible with an invertible state, but the allowed invertible phase is shifted relative to that of an onsite $\Z_4^F$ symmetry.

Moreover, stacking this $p+ip$ state with an arbitrary onsite $\Z_4^F$-symmetric invertible phase of $c_-\in \Z$ does not change the anomaly index $[n_2]$, while shifting the chiral central charge by $k$. We therefore obtain symmetric invertible states with
$c_-
=
k+1/2$ with $c_-\in \Z$.

\subsubsection{Onsite $\Z_4^F$ symmetry: Invertible states with $c_-\in \Z$}

We also argue that invertible states with an onsite $\Z_4^F$ symmetry must have integer chiral central charge $c_- \in \Z$.
We will show this by contradiction.
To this end, we suppose that an onsite $\Z_4^F$ symmetry admits an invertible state $\ket{\Psi}$ with $c_- \in \Z+1/2$.
By stacking $\ket{\Psi}$ with the $p+ip$ state $\ket{p+ip}$, we obtain an invertible state $\ket{\Psi} \otimes \ket{p+ip}$ with $c_- \in \Z$.
This state has a diagonal non-onsite $\Z_4^F$ symmetry with a non-trivial lattice anomaly $[n_2] \neq 0$.
However, such a non-onsite $\Z_4^F$ symmetry is incompatible with $c_- \in \Z$, as discussed in the previous subsection.
This implies that there does not exist an invertible state $\ket{\Psi}$ with the above properties.
Thus, we find that an onsite $\Z_4^F$ symmetry is incompatible with an invertible state with $c_- \in \Z+1/2$.

\section{$\Z_4^F$ symmetry in (1+1)D}
\label{app:Z4 in 1+1D}
 Let us consider a chain with a single complex fermion $c_j$ per site, represented by a pair of Majorana fermions $a_j, b_j$. We consider the $G_f=\Z_4^F$ symmetry with the extension $\Z_2^F\to G_f\to G_b$ with $G_b=\Z_2$.
 
The $\Z_4^F$ symmetry is generated by a finite-depth circuit $U$ transforming the Majorana fermions as
\begin{align}
    \quad b_j\to a_{j+1}~, \ a_{j+1}\to -b_{j}~.
\end{align}
One can immediately see $U^2=(-1)^F$. 

One can see that the anomaly index $[n_2]\in H^2(BG_b,\Z_2)$ associated with this symmetry operator is nontrivial, therefore the symmetry forbids an SRE state~\cite{else2014, Seifnashri2024lsm, chavda2026fermionicanomalies}.
The anomaly index is defined as follows: when the symmetry is generated by a finite-depth circuit $U(g)$ labeled by $g\in G_b$, we first truncate the symmetry operator to the interval $I$, then evaluate the projective action
\begin{align}
    U_I(g)U_I(h) = \Omega_{\partial I}(g,h)[(-1)^{F_I}]^{\omega_2(g,h)}U_I(gh)~,
\end{align}
then $\Omega_{\partial I}(g,h)$ is a product of endpoint operators at $\partial I$ and has the form of $\Omega_{\partial I} = \Omega_l\Omega_r$ at the left and right ends. The fermion parity of $\Omega_l(g,h)$ defines the $\Z_2$-valued cocycle $n_2(g,h)\in Z^2(BG_b,\Z_2)$, and its cohomology class obstructs SRE.

For the $\Z_4^F$ symmetry defined above, one can see $U_I^2 \propto a_{l} b_r(-1)^{F_I}$, where $a_l, b_r$ are the Majorana fermions at the left, right ends of the interval. The above algebra involving boundary Majorana operators $a_l, b_r$ corresponds to the nontrivial index $[n_2]\in H^2(BG_b,\Z_2)$. This index obstructs SRE.

Nevertheless, the above $\Z_4^F$ symmetry still preserves an invertible state which is not SRE. The invertible state is the Kitaev's Majorana chain which is a ground state of the Hamiltonian
\begin{align}
    H=-\sum_{j}ib_ja_{j+1}~.
\end{align}
This preserves the $\Z_4^F$ symmetry generated by $U$. 

The classification of invertible phases with this $\Z_4^F$ symmetry is trivial, identical to the case of onsite $\Z_4^F$ symmetry. This is because one can transform the symmetry $U$ into an onsite form by a Majorana translation operator $\mathcal{T}$: $\mathcal{T}^{-1}U\mathcal{T}$ is an onsite $\Z_4^F$ symmetry. Therefore the Majorana translation induces an isomorphism between the invertible phase classifications with onsite and non-onsite symmetry. Thus, the above Kitaev's Majorana chain is the single invertible phase with the non-onsite $\Z_4^F$ symmetry carrying the $[n_2]\in H^2(BG_b,\Z_2)$ index. 
We note that this (1+1)D Kitaev chain with $\mathbb Z_4^F$ symmetry was discussed in \cite{WangGu2020} as a boundary theory of a (2+1)D $\mathbb Z_4^F$ SPT phase. Such a boundary state of a trivial bulk SPT phase was referred to as an anomalous SPT state.

\bibliographystyle{utphys}
\bibliography{bibliography}

\end{document}